\documentclass[11pt,twoside]{article}
\usepackage{asp2014}

\aspSuppressVolSlug
\resetcounters

\begin{document}

\title{Integration of storage endpoints into a Rucio data lake, as an activity to prototype a SKA Regional Centres Network}

\author{Manuel~Parra-Royón$^1$, Jesús~Sánchez-Castañeda$^1$, Julián~Garrido$^1$, Susana Sánchez-Expósito$^1$, Rohini~Joshi$^2$, James~Collinson$^2$, Rob~Barnsley$^2$, Jesús~Salgado$^2$ and Lourdes~Verdes-Montenegro$^1$}
\affil{$^1$ Instituto de Astrofísica de Andalucía - CSIC, Granada, Spain}
\affil{$^2$ Square Kilometer Array Observatory - SKAO, United Kingdom}

\paperauthor{Manuel~Parra-Royón}{mparra@iaa.es}{ORCID}{0000-0002-6275-8242}{Extragactic Astronomy}{Granada}{Granada}{18008}{Spain}
\paperauthor{Jesús~Sánchez-Castañeda}{jsanchez@iaa.es}{ORCID}{}{Extragactic Astronomy}{Granada}{Granada}{18008}{Spain}
\paperauthor{Julián~Garrido}{jgarrido@iaa.es}{ORCID}{}{Extragactic Astronomy}{Granada}{Granada}{18008}{Spain}
\paperauthor{Susana~Sánchez-Expósito}{sse@iaa.es}{ORCID}{}{Extragactic Astronomy}{Granada}{Granada}{18008}{Spain}
\paperauthor{Rohini~Joshi}{rohini.joshi@skao.int}{ORCID}{}{SKAO}{}{}{}{UK}
\paperauthor{James~Collinson}{james.collinson@skao.int}{ORCID}{}{SKAO}{}{}{}{UK}
\paperauthor{Rob~Barnsley}{rob.barnsley@skao.int}{ORCID}{}{SKAO}{}{}{}{UK}
\paperauthor{Jesús~Salgado}{jesús.salgado@skao.int}{ORCID}{}{SKAO}{}{}{}{UK}
\paperauthor{Lourdes~Verdes-Montenegro}{lourdes@iaa.es}{ORCID}{0000-0003-0156-6180}{SKAO}{}{}{}{UK}




  
\begin{abstract}

The Square Kilometre Array (SKA) infrastructure will consist of two radio telescopes that will be the most sensitive telescopes on Earth. The SKA community will have to process and manage near exascale data, which will be a technical challenge for the coming years. In this respect, the SKA Global Network of Regional Centres plays a key role in data distribution and management. The SRCNet will provide distributed computing and data storage capacity, as well as other important services for the network. Within the SRCNet, several teams have been set up for the research, design and development of 5 prototypes. One of these prototypes is related to data management and distribution, where a data lake has been deployed using Rucio. In this paper we focus on the tasks performed by several of the teams to deploy new storage endpoints within the SKAO data lake. In particular, we will describe the steps and deployment instructions for the services required to provide the Rucio data lake with a new Rucio Storage Element based on StoRM and WebDAV within the Spanish SRC prototype.

\end{abstract}

\section{Introduction}

The Square Kilometre Array Observatory (SKAO) will build the most sensitive telescope on the planet to address key questions in astrophysics, fundamental physics and astrobiology. SKA telescope will consist of two radio telescopes, one in Australia housing 131,072 'Christmas tree' antennas, grouped into 512 stations, each with 256 antennas, and another in South Africa with an array of 197 antennas. In the process of developing this major scientific infrastructure, precursors and pathfinders, such as MeerKAT and HERA have been put in place together with the Murchison Widefield Array (MWA) and the CSIRO's Australian SKA Pathfinder (ASKAP), which provide SKA scientists and engineers with the knowledge and technology needed for the future design and deployment of the SKA. Both these facilities and the future SKA main telescope will require the capacity to process and manage a huge amount of data, close to Exascale which will be a technical challenge for the coming years given its volume, velocity, variety, veracity, and value. To allow the storage and analysis of data at this scale, data must be distributed within the SRC Network (SRCNet) nodes, pledging, and sharing resources internationally. 
The SRC Steering Committee (SRCSC) is in charge of defining and creating an operational partnership between SKAO and a set of independently resourced SRCs, pledging, and sharing storage and computational resources to the SRCNet. In this context, and as part of the work to be developed within the different working groups involved, 5 prototypes have been proposed and are being addressed with the aim of implementing each of the different building blocks that will shape the future SRCNet. The prototypes bring together the key aspects for science delivery such as storage distribution and management, authentication infrastructure, data visualisation, science platform and software distribution and delivery. 

In this paper we will focus on the work that has been developed on the first prototype for storage management and distribution and specifically on the process of integrating a new Rucio Storage Element (RSE) from the Spanish SRC (SPSRC) into an existing Rucio-based data lake prototype at SKAO.

\section{Context}

The work for each of the SRCNet prototypes involves a huge planning, organisation and development effort that necessarily has to be governed effectively. To this end, an SRC Agile Release Train (SRC ART) has been set up within the SRCSC, following a Scaled Agile methodology, to set-up different kinds of teams (stream-aligned, complicated subsystem, platform orenabling teams among others) and work teams for the development of features within these 5 prototypes. Cyan team, a platform team is in charge of the storage management prototype, Orange team is tackling the visualisation tools, Tangerine team (platform team) is working to define the science platform and other teams such as Coral and the Blue-Lavender team are stream-aligned, responsible to build and deliver customer value and work on the flow with the rest. In this context Cyan Team has provided a functional prototype testbed with a Rucio-based data lake for data distribution, and other teams like Coral team and Blue-Lavender have worked on the integration of new RSEs, one in Spain \cite{10.1117/1.JATIS.8.1.011004} and the another in Japan (JPSRC), with the aim of extending the global distributed data storage network to other SRCs. To proceed with this integration work, coral and cyan team have been working on the deployment of the storage infrastructure, deployment and configuration of services for the addition of a new RSE for the prototype SKAO data lake. As a result, several documentation sources have been created to describe this integration, as well as two repositories to simplify and automate the process of deploying a new RSE within the Rucio data lake.

\section{Deployment}

\begin{figure}[ht]
\includegraphics[width=0.8\textwidth]{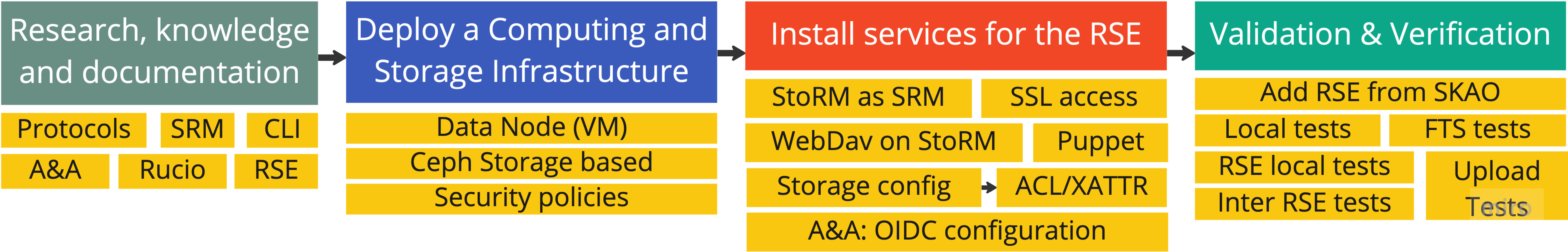}
\centering
\label{fig:steps}
\caption{Phases for the integration of the SPSRC RSE.}
\end{figure}

The work of deploying a new RSE for the Rucio data lake has consisted of 4 phases, as described in Figure \ref{fig:steps}. A first phase consisting of research and documentation in which, in coordination with the Cyan team, we reviewed the available information, storage management technologies, access protocols and services to deploy an RSE. Following this was the deployment of the computing and storage infrastructure, Virtual Machine (VM) and Ceph Storage that was used to create a new endpoint in the data lake. The third phase was the installation of the services and software, as well as the configuration of the storage manager and the services to specify the connection protocol between the RSE and FTS. Finally, we added the RSE to the Rucio SKAO data lake prototype and verified the rest of the RSEs.

\subsection{Study and knowledge acquisition}

Rucio has the ability to integrate storage elements with a huge variety of connection protocols, which makes it highly flexible, while at the same time homogenising data access/management even if the RSEs have different protocols. Key to this is the File Transfer System (FTS), which plays an important role in enabling network interoperability between endpoints. 
For this purpose, a study of the existing RSEs, their connection protocols and the features of the services providing storage management was carried out. Within the SKAO Rucio prototype data lake, several storage resource managers (SRMs) such as EOS, dCache or StoRM\footnote{\url{https://italiangrid.github.io/storm/index.htm}} and protocols such as WebDAV or xRoot have already been tested. For our case study we chose to install StoRM as the SRM and use WebDAV over StoRM as the RSE connection protocol, the decisive factor being the availability of  documentation, support and capabilities for OIDC token Authentication and Authorisation (A\&A).

\subsection{RSE infrastructure}

The StoRM SRM and transfer protocol services (WebDAV) were installed on computing and storage infrastructure. For the deployment at the SPSRC we have created a VM, running CentOS 8 and with 4 vCPUs, 8GB of RAM and a 10TB attached storage based on Ceph Shared Filesystem (CephFS), so it is possible to dynamically extend or reduce the space delivered to the data lake on demand. 

\subsection{StoRM, WebDAV deployment and A\&A configuration}

As discussed in the previous sections, for the RSE it was chosen to install StoRM as the SRM and WebDAV as the transfer protocol for connection with the rest of the RSEs/FTS and the data lake. The manual installation was done using the instructions that were compiled in the study phase that provided all the basic configuration for both the SRM and the WebDAV service in StoRM. For this manual installation a repository was created\footnote{\url{https://github.com/spsrc/rucio-rse-StoRM-webdav}} that contains in detail all the steps from configuration of the VM, the installation of the services, the preparation of the A\&A for the RSE and the configuration of the shared storage. In the first part the Ceph-based shared storage and the attributes for this type of storage are configured and the StoRM repositories and the basic services are installed. The SRM is then configured with the storage parameters, the external access point, size and the supported protocols for access, such as \texttt{WebDAV}, \texttt{file} or \texttt{gsiftp}. Finally, the OIDC-based A\&A is configured using ESCAPE IAM\footnote{\url{https://iam-escape.cloud.cnaf.infn.it}} as the A\&A provider and adding a specific client for it. Once these tasks were completed, we restarted the services and verified that they were running correctly in the RSE. Additionally, a repository\footnote{\url{https://github.com/spsrc/RSE-Ansible}} was developed to tackle the automation of the entire installation process using Ansible.

\section{Integration and validation within SKAO prototype data lake}

Once the services were up and running and are externally exposed, it was necessary to include the configuration and parameterisation of the RSE within Rucio data lake. To do so, Cyan team, the team responsible for the development and deployment of the data lake in SKAO, was contacted to request the addition of the SPSRC RSE using administrative processes. This operation is simple and requires only the basic data of the previously configured service; in our specific case, the name: \texttt{SPSRC\_STORM}, the endpoint with webdav for SPSRC: \texttt{https://spsrc14.iaa.csic.es:18027}, and the Rucio prefix: \texttt{/disk/storm/sa/}. The final step after integration was to verify that all blocks were connected, from connection and access to the RSE itself using \texttt{davix} (with A\&A based token configuration), access via FTS with \texttt{fts-rest-*} tools and copying between and to RSEs using the Rucio CLI.

\section{Conclusions and future work}

Rucio is an outstanding platform for the distribution/management of massive data and offers a wide set of protocols for the integration of new RSEs. In this work we focus on an SRM, StoRM, and specific protocol, WebDAV, as well as a CephFS-based storage, demonstrating that this integration is possible within the SKAO data lake and providing 2 detailed installation modes for implementation by other SRCs. As future work, aspects such as the integration of full object-based storage, as well as testing the limits of the platform globally, and the integration of the data lake with the computational resources for the analysis of the scientific data, remain to be solved.

\section*{Acknowledgments}
We acknowledge financial support from the State Agency for Research of the Spanish Ministry of Science, Innovation and Universities through the “Center of Excellence Severo Ochoa (SEV-2017-0709) and RTI2018-096228-B-C31, and the financial support from the Grant No. 54A Scientific Research, Innovation Program (Government of Andalusia and the European RDF 2014-2020 - D1113102E3). We also acknowledge the financial support from the grant PID2021-123930OB-C21 funded by MICIU/AEI/ 10.13039/501100011033 and by ERDF/EU.

\bibliography{P62}

\end{document}